\RequirePackage{lineno}
\documentclass[prl,twocolumn,showpacs,amsmath,amssymb]{revtex4-1}

\usepackage{graphicx}
\usepackage{dcolumn}
\usepackage{bm}
\usepackage{rotating}
\usepackage{epstopdf}
\usepackage{color}
\usepackage{verbatim} 
\usepackage{multirow}
\usepackage[abs]{overpic}
\usepackage{amsmath}
\usepackage{mathrsfs}
\usepackage{amssymb}
\usepackage{xspace}
\usepackage{float}
\usepackage[caption=false]{subfig}

\usepackage[colorlinks,
            linkcolor=blue,
            anchorcolor=blue,
            citecolor=blue]{hyperref}

\newcommand{\PreserveBackslash}[1]{\let\temp=\\#1\let\\=\temp}
\newcolumntype{C}[1]{>{\PreserveBackslash\centering}p{#1}}
\newcolumntype{R}[1]{>{\PreserveBackslash\raggedleft}p{#1}}
\newcolumntype{L}[1]{>{\PreserveBackslash\raggedright}p{#1}}

\begin{document}
\graphicspath{{figure/}}
\DeclareGraphicsExtensions{.eps,.png,.ps}
\title{\boldmath High-precision measurement of the space-like $\eta^\prime$ transition form factor }

\author{
\begin{small}
\begin{center}
M.~Ablikim$^{1}$\BESIIIorcid{0000-0002-3935-619X},
M.~N.~Achasov$^{4,c}$\BESIIIorcid{0000-0002-9400-8622},
P.~Adlarson$^{84}$\BESIIIorcid{0000-0001-6280-3851},
X.~C.~Ai$^{90}$\BESIIIorcid{0000-0003-3856-2415},
C.~S.~Akondi$^{32A,32B}$\BESIIIorcid{0000-0001-6303-5217},
R.~Aliberti$^{40}$\BESIIIorcid{0000-0003-3500-4012},
A.~Amoroso$^{83A,83C}$\BESIIIorcid{0000-0002-3095-8610},
Q.~An$^{79,66,\dagger}$,
Y.~H.~An$^{90}$\BESIIIorcid{0009-0008-3419-0849},
M.~S.~Anderson$^{40}$\BESIIIorcid{0009-0008-1550-2632}
% ...
\\
\vspace{0.2cm}
(BESIII Collaboration)\\
\vspace{0.2cm} {\it
$^{1}$ Institute of High Energy Physics, Beijing 100049, People's Republic of China\\
% ...
$^{4}$ Budker Institute of Nuclear Physics SB RAS (BINP), Novosibirsk 630090, Russia\\
% ...
$^{32}$ INFN Sezione di Ferrara, (A)INFN Sezione di Ferrara, I-44122, Ferrara, Italy; (B)University of Ferrara, I-44122, Ferrara, Italy\\
% ...
$^{40}$ Johannes Gutenberg University of Mainz, Johann-Joachim-Becher-Weg 45, D-55099 Mainz, Germany\\
% ...
$^{66}$ State Key Laboratory of Particle Detection and Electronics, Beijing 100049, Hefei 230026, People's Republic of China\\
% ...
$^{79}$ University of Science and Technology of China, Hefei 230026, People's Republic of China\\
% ...
$^{83}$ University of Turin and INFN, (A)University of Turin, I-10125, Turin, Italy; (B)University of Eastern Piedmont, I-15121, Alessandria, Italy; (C)INFN, I-10125, Turin, Italy\\
$^{84}$ Uppsala University, Box 516, SE-75120 Uppsala, Sweden\\
% ...
$^{90}$ Zhengzhou University, Zhengzhou 450001, People's Republic of China\\
\vspace{0.2cm}
$^{\dagger}$ Deceased\\
% ...
$^{c}$ Also at the Novosibirsk State University, Novosibirsk, 630090, Russia\\
% ...
}
\end{center}
\vspace{0.4cm}    
\end{small}
}

\vspace{0.2cm}
\date{\today}

\begin{abstract}
Using a data sample corresponding to an integrated luminosity of $20.3\ \text{fb}^{-1}$, collected with the BESIII detector at a center-of-mass energy of $3.773\ \text{GeV}$ at the BEPCII collider, 
we report a precision measurement of the product $Q^2|F(Q^2)|$, where $F(Q^2)$ is the single-virtual space-like transition form factor of the $\eta'$ meson and $Q^2$ is the squared momentum transfer of the tagged virtual photon.
The transition form factor is extracted from the differential Born cross section of the two-photon fusion processes $e^+e^- \to e^+e^-\gamma\gamma^* \to e^+e^-\eta^\prime$ using a single-tag technique, where only one scattered lepton is detected. 
The measurement covers $Q^2 \in [0.1, 6.0]$ GeV$^2$, achieving unprecedented precision, better than $3.0\%$ for $Q^2 < 1.5$ GeV$^2$, and providing the first direct determination at $Q^2 < 0.3$ GeV$^2$.
\end{abstract}

\newcommand{\BESIIIorcid}[1]{\href{https://orcid.org/#1}{\hspace*{0.1em}\raisebox{-0.45ex}{\includegraphics[width=1em]{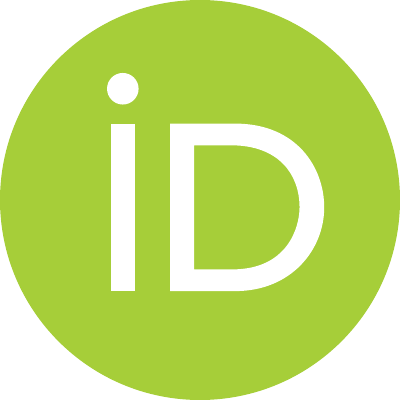}}}}

\maketitle

The photon-meson transition form factor (TFF) characterizes the coupling of a meson to two (virtual)
photons, $\gamma^{(*)}\gamma^{(*)}\to M$, and provides a key probe of both the internal structure of mesons and low-energy quantum chromodynamics (QCD). 
The space-like TFF, $F(q_1^2, q_2^2)$, depends on the photon virtualities $q_i^2$ ($i=1,2$),
and can be accessed through the two-photon fusion processes 
\begin{equation*}
e^+(p_1)e^-(p_2) \to e^+(p'_1) e^-(p'_2) \gamma^{(*)}(q_1)\gamma^{(*)}(q_2)
\to e^+ e^- M,
\end{equation*}
where $p_i$ and $p'_i$ denote the four-momenta of the initial-
and final-state leptons, and the respective momentum transfers
$q_i^2 \equiv -Q_i^2 = (p_i-p'_i)^2$ denote the four-momenta of the intermediate photons.
The TFF can also be studied in the time-like region through the meson Dalitz 
decays or in electron-positron annihilation~\cite{TFF_HLBL}.
Measurements of the TFF at small $Q^2$ are essential for determining 
the charge radii of mesons, while those at large $Q^2$ test the meson distribution 
amplitudes predicted by perturbative QCD~\cite{DAs}. 

For the pseudoscalar mesons, $\pi^0$, $\eta$, and $\eta^\prime$, the TFFs 
are direct inputs for the calculation of hadronic light-by-light (HLbL) 
contribution to the muon anomalous magnetic moment $a_{\mu}=(g-2)/2$~\cite{white_paper, white_paper2020} 
in data-driven approaches~\cite{DataDriven1_pi0, DataDriven2_EtaEtap, DataDriven3_EtaEtap}. 
They also provide essential benchmarks for lattice QCD (LQCD)~\cite{LQCD2020_Eta, LQCD2020_Pi0EtaEtap, LQCD2020_3, LQCD2020_4} 
or holographic model calculations~\cite{Holographic}.
In the HLbL evaluation, recent LQCD calculations yield a value about 50\% higher than the 2020 white paper estimate, which further amplifies the tension between LQCD and data-driven approaches~\cite{white_paper, white_paper2020}.
Furthermore, the $\eta^\prime$ TFF can be related to the $\eta^\prime$-gluon-gluon
coupling, which is sensitive to the gluonic component of the $\eta^\prime$
wave function~\cite{HHC}. This coupling also governs $\eta^\prime$ production mechanisms 
in hadronic and heavy-ion collisions, offering a unique probe of the 
nonperturbative QCD dynamics.

\begin{figure*}[htp]  
\centering
\includegraphics[width=0.43\textwidth]{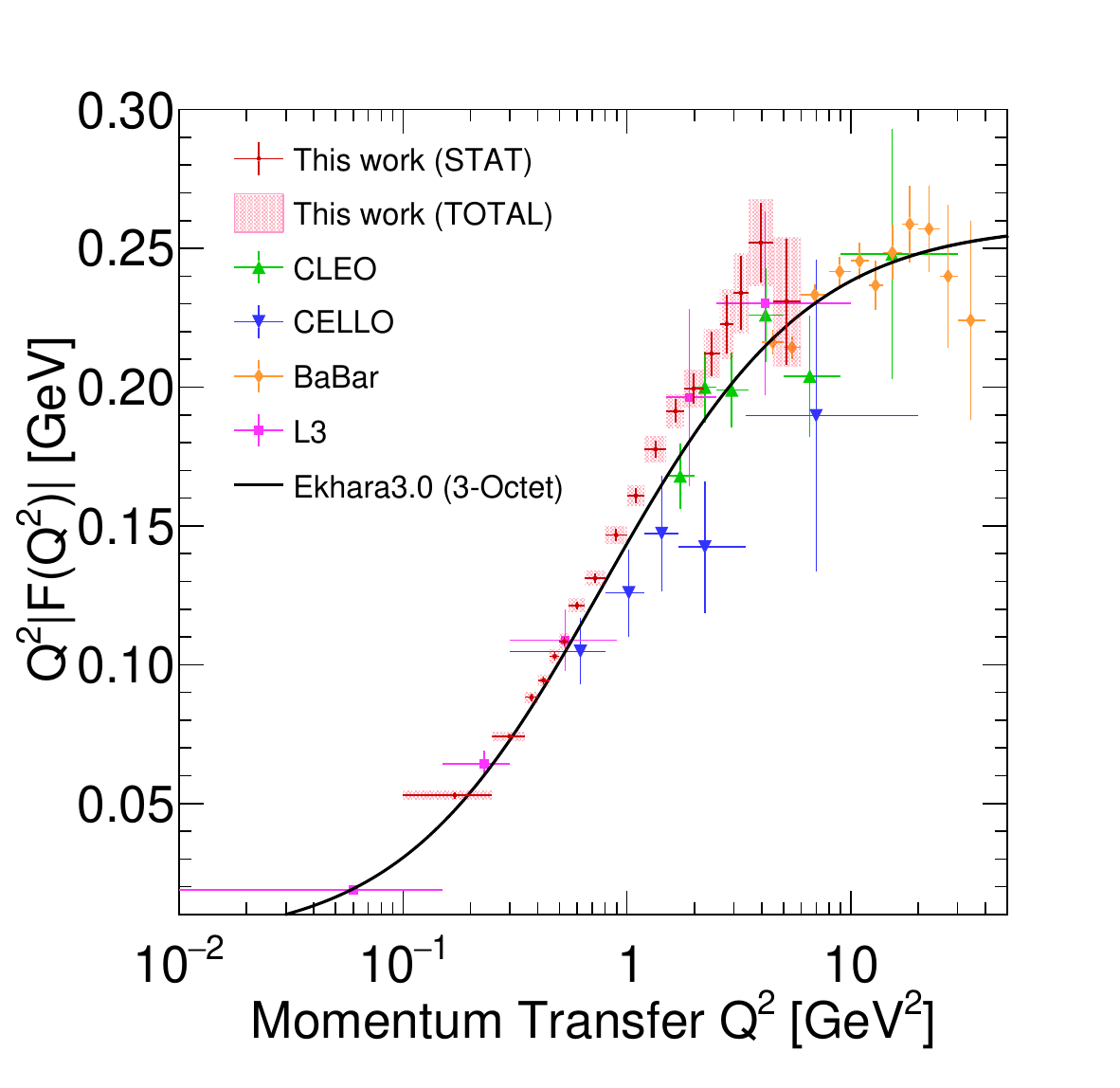}
\includegraphics[width=0.43\textwidth]{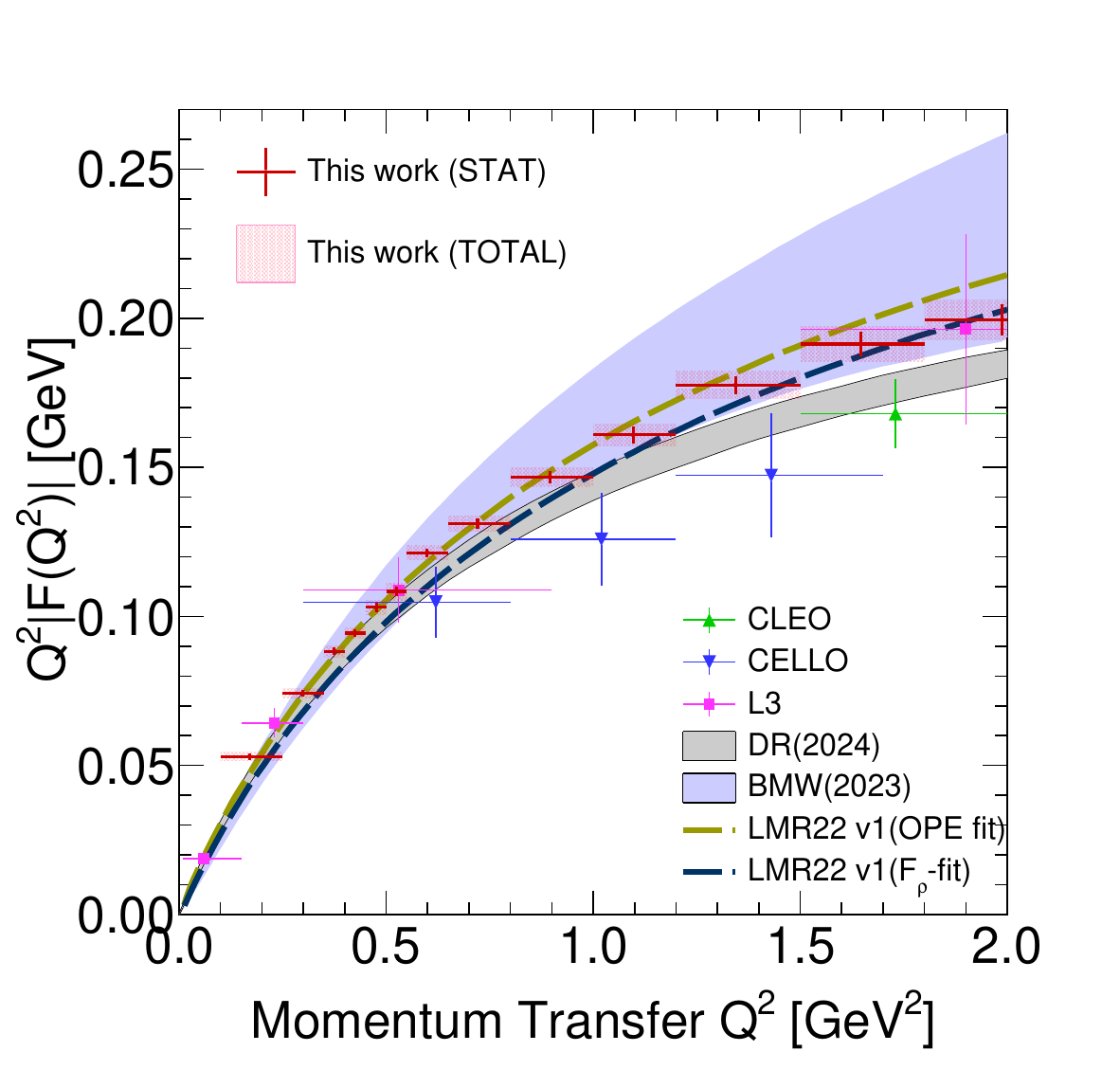}
\caption{
Space-like $\eta^{\prime}$ TFF in the single-tag configuration, compared with the results from CELLO~\cite{ICELLO}, CLEO~\cite{ICLEO}, L3~\cite{IL3}, BaBar~\cite{IBABAR}, and model implemented in the {\sc ekhara} generator~\cite{EKHARA} (left). Space-like $\eta^\prime$ TFF in the region $Q^2<2.0~\mathrm{GeV}^2$, compared with theoretical calculations from 
data-driven (DR)~\cite{DataDriven3_EtaEtap},
LQCD (BMW)~\cite{LQCD2020_Pi0EtaEtap}, and 
Holographic (LMR)~\cite{Holographic} approaches (right). 
Red dots with error bars (pink shaded boxes) indicate statistical (total) uncertainties of this work.
}
\label{fig:sys_all}
\end{figure*}

The space-like TFFs of pseudoscalar mesons have been measured by several experiments, including CELLO~\cite{ICELLO}, CLEO~\cite{ICLEO}, L3~\cite{IL3}, Belle~\cite{IBELLE}, BaBar~\cite{IBABAR_pi0, IBABAR, IBABAR_doubletag}, and BESIII~\cite{IBESIII}. 
Existing measurements of the single-virtual $\eta^{\prime}$ transition form factor, corresponding to $F(Q^2, Q_{\rm miss}^2 \approx 0) \equiv F(Q^2)$
and obtained with the single-tag technique, are summarized in Fig.~\ref{fig:sys_all}, 
In this approach, one final-state lepton is detected (tagged), while the other is scattered outside the detector acceptance, 
leaving the corresponding photon quasi-real $(Q_{\rm miss}^2 \approx 0)$.
Precise measurements exist mainly in the high-$Q^2$ region ($Q^2>4.0~\mathrm{GeV}^2$),
while the low-$Q^2$ domain (${Q}^{2} < 1.5~\mathrm{GeV}^{2}$), crucial for the HLbL 
evaluation~\cite{ANyffeler}, remains poorly constrained. 
Notably, the L3 result for $Q^2<1.0~\mathrm{GeV}^2$ infers the $Q^2$ 
dependence indirectly from the $\eta^\prime$ transverse momentum distribution. 
A precise measurement of the $\eta'$ TFF at low $Q^2$ is therefore essential for improving the HLbL calculation of $a_\mu$, for probing the gluonic content of the $\eta'$, and for constraining models of $\eta'$ production.


In this Letter, we present a precise measurement of the single-virtual TFF of 
the $\eta^{\prime}$ meson in the space-like region, covering the momentum-transfer 
range $Q^{2}\in [0.1,~6.0]~\mathrm{GeV}^2$.
The analysis is based on a data sample 
corresponding to an integrated luminosity of $20.3~\mathrm{fb}^{-1}$~\cite{DATA0, DATA}, 
collected with the BESIII detector~\cite{Ablikim:2009aa} at the BEPCII 
collider operating~\cite{Yu:IPAC2016-TUYA01} at a center-of-mass energy of 
$\sqrt{s}=3.773~\mathrm{GeV}$. The process under study is $e^+e^- \to e^+e^-\gamma\gamma^* \to e^+e^-\eta^\prime$,
where the $\eta^{\prime}$ is reconstructed in its decay modes: $\pi^+\pi^-\gamma$ (mode I)
and $\pi^+\pi^-\eta$ with $\eta\to\gamma\gamma$ (mode II).

Monte Carlo (MC) simulated data samples produced with a 
{\sc geant4}-based~\cite{GEANT4} software package, 
which includes the geometric description of the BESIII detector and 
the detector response, are used to determine detection efficiencies and 
to estimate backgrounds. 
The simulation models the beam energy spread and initial-state radiation 
(ISR) in the $e^+e^-$ annihilations with the generator {\sc kkmc}~\cite{KKMC}. 
The inclusive MC sample includes the production of $D\bar{D}$ pairs 
(including quantum coherence for the neutral $D$ channels), the non-$D\bar{D}$ 
decays of the $\psi(3770)$, the ISR production of the $J/\psi$ and $\psi(3686)$ 
states, and the continuum processes incorporated in {\sc kkmc}~\cite{KKMC}.
All particle decays are modeled with {\sc evtgen}~\cite{BesEvtGen} using 
branching fractions either taken from the Particle Data Group (PDG)~\cite{PDG}, 
when available, or otherwise estimated with {\sc lundcharm}~\cite{lundcharm}.
Final state radiation from charged final state particles is incorporated 
using the {\sc photos} package~\cite{PHOTOS}.

Signal $e^+e^- \rightarrow e^+e^-\gamma \gamma^{\ast} \rightarrow e^+e^-\eta^{\prime}$ events 
are generated with {\sc ekhara3.0} generator~\cite{EKHARA3.0}. The interaction of photons, 
pseudoscalars and vector mesons are modeled within the resonance chiral
theory with SU(3) breaking, and full next-to-leading-order radiative
corrections are included. 
The TFF is modeled with a triple-octet framework with parameters determined from 
a fit to the available data in both space-like and time-like regions~\cite{EKHARA}. 
Mode I and mode II are modeled in {\sc evtgen} according to recent precision data on the decay dynamics~\cite{Etap2gpipieta, Etap2gpipi}.
Dedicated MC samples are produced for the background processes, 
$e^+e^-\to e^+e^- \eta$, $e^+e^-\to e^+ e^- \pi^+ \pi^-$, and 
$e^+e^- \to e^+e^- \mu^+\mu^-$ using {\sc ekhara}~\cite{EKHARA}, {\sc diag36},
and {\sc bdkrc} event generators~\cite{diag36, bdkrc}. Radiative corrections 
are included for $e^+e^-\to e^+e^- \eta$ events and $e^+e^-\to e^+e^- \mu^+\mu^-$ 
events through the two-photon process. 


Charged tracks reconstructed in the multilayer drift chamber (MDC) are required 
to be within a polar angle ($\theta$) range of $|\cos\theta| < 0.93$, where 
$\theta$ is defined with respect to the $z$-axis, which is the symmetry axis 
of the MDC. The distance of closest approach to the interaction point (IP) 
must be less than 10 cm along the $z$-axis and less than 1 cm in the transverse
plane. 
Exactly three charged tracks are required with a total charge of $\pm1$ 
matching the charge of the tagged lepton.

Particle identification (PID) combines measurements of the 
specific ionization energy loss
in the MDC ($dE/dx$) and the time-of-flight system to form likelihoods $\mathcal{L}(i)$ ($i = \pi, K, p, e$) 
for each hypothesis. A track is identified as a pion if $\mathcal{L}(\pi) > 0.001$
and $\mathcal{L}(\pi)$ is the largest among the tested hypotheses. The event 
must contain exactly one positively and one negatively charged pion. 
The remaining track is identified as $e^{\pm}$ if the ratio $E/p$
of its deposited energy in the electromagnetic calorimeter (EMC) to 
its momentum measured in the MDC exceeds 0.8. Exactly one electron or 
positron candidate is required.

Photon candidates are identified using showers in the EMC.
The deposited energy of each shower must be more than 25 MeV in the barrel 
region ($|\cos\theta|<0.8$) and more than 50 MeV in the end-cap regions 
($0.86<|\cos\theta|<0.92$).
To exclude showers that originate from charged tracks, the angle subtended 
by the EMC shower and the position of the closest charged
track at the EMC must be greater than 10 degrees as measured from the IP.
To suppress electronic noise and showers unrelated to the event, 
the difference between the EMC shower time and the event start time is 
required to be within $[0,~700]~\mathrm{ns}$.
The number of accepted photon candidates must be 
$\geq$ 1 for mode I and $\geq$ 2 for mode II.
For mode II, the $\eta$ meson candidates are reconstructed from pairs of photons.
A one-constraint 
(1C) kinematic fit is performed by constraining the invariant mass $M_{\gamma\gamma}$ 
of the two photons to the nominal $\eta$ mass~\cite{PDG}. Combinations 
with $\chi^2_{\mathrm{1C}\text{-}\eta} < 200$ are retained for further analysis. 

To suppress background and improve mass resolution, a kinematic 
fit is performed to the global event. In mode I, a 1C kinematic fit is applied by constraining 
the four-momentum of the final-state particles 
$\gamma\pi^{+}\pi^{-}e^{\pm}e^{\mp}_{\rm miss}$ to the initial four-momenta, 
where $e_{\rm miss}$ denotes the missing track with its mass fixed to the 
electron/positron hypothesis. In mode II, a two-constraint (2C) kinematic 
fit is employed, in which the invariant mass of the photon pair is additionally 
constrained to the nominal $\eta$ mass. 
The best $\gamma$ candidate in mode I and the best $\eta$ candidate in mode II are 
selected according to the minimum $\chi^{2}_{\mathrm{1C}}$ and 
$\chi^{2}_{\mathrm{2C}}$, respectively. 
Events are required to satisfy $\chi^2_{\rm 1C}<70$ in mode I and 
$\chi^2_{\rm 2C}<200$ in mode II, where the requirements are optimized 
according to $S/\sqrt{S+B}$, with $S$ and $B$ denoting 
the number of signal and background events estimated from  MC simulations. 
The four-momenta from the 1C (mode I) or 2C (mode II) fits are used in
subsequent calculations.

To ensure small momentum transfer of the quasi-real photon, 
the scattering angle $\theta_{\rm miss}$ of the untagged lepton 
is required to satisfy $\cos\theta_{\mathrm{miss}} > 0.99$. 
After this requirement, the $Q^{2}$ associated with the untagged lepton, 
$Q^{2}_{\rm miss}$, is limited to 
the range of $[0,~0.08]~\mathrm{GeV}^2$ according to signal MC,
and the distribution is shown in the supplemental material~\cite{supp}.

\begin{figure}[htp]
\centering
\includegraphics[scale=0.4]{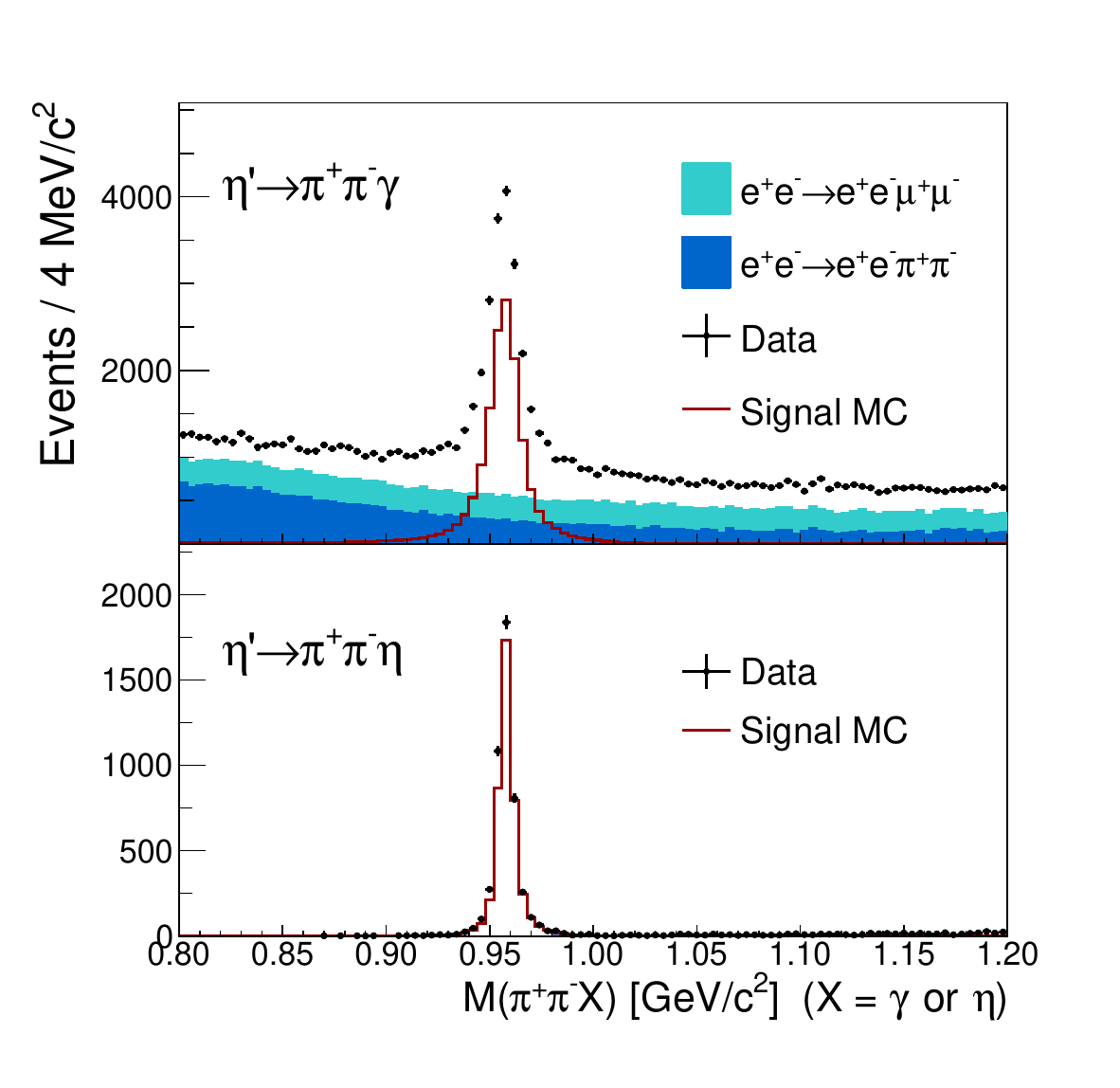}
\includegraphics[scale=0.4]{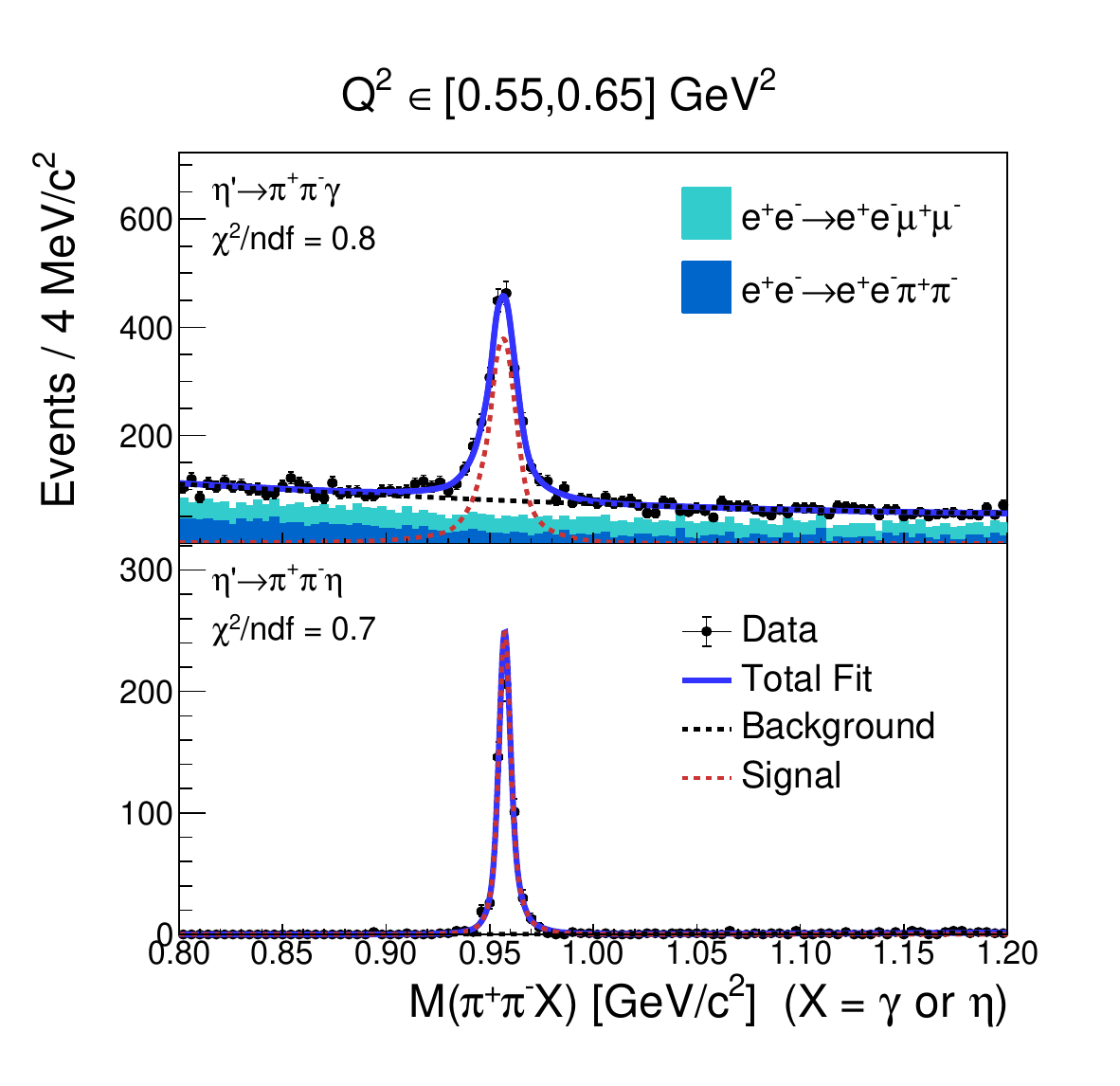}
\caption{The $\pi^+\pi^-\gamma$ and $\pi^+\pi^-\eta$ invariant mass 
distributions summed over all $Q^2$ bins (upper histogram), together with the simultaneous
fit result for $Q^2\in [0.55,~0.65]~\mathrm{GeV}^2$ (lower histogram). 
The dots with error bars are data, the red histograms correspond to the signal MC and the shaded histograms show the stacked background contributions 
estimated from MC simulations.
The blue solid curves denote the total fit results, the red dashed lines represent the signal processes, 
and the black dashed lines indicate the background contributions. 
}
\label{fig:com_all}
\end{figure}

In mode I, the dominant backgrounds arise from the processes $e^+e^-\to e^+e^-l^+l^-$ 
and $e^+e^-\to e^+e^-\pi^+\pi^-$. In the case of $l=\mu$, misidentified 
muons contribute as pions, and the resulting pion pair is combined with 
a fake photon to mimic the signal.
The corresponding $\pi^+\pi^-\gamma$ invariant mass distributions are smooth. 
Background contributions from other processes are found to be negligible. 
In mode II, the background is negligible because the selection requires 
the $\eta$ meson to be reconstructed.
Figure~\ref{fig:com_all} shows the $\pi^+\pi^-\gamma$ or $\pi^+\pi^-\eta$ 
invariant mass distributions for the selected $\eta^\prime$ candidate events, 
together with the estimated stacked $e^+e^-\mu^+\mu^-$ and $e^+e^-\pi^+\pi^-$ 
backgrounds. 

To investigate the $\eta^\prime$ TFF as a function of $Q^2$, the signal
yield in each $Q^{2}$ interval ($Q_{\mathrm{bin},i}^2$) is determined by an unbinned 
simultaneous maximum likelihood fit to the $\eta^\prime$ invariant mass spectrum from 
mode I and mode II within $0.8$ to $1.2~\mathrm{GeV}/c^2$, including both signal and background components.
The common parameter in the simultaneous fit is the differential cross section.
The selected events cover the region $0.1\leq Q^2 \leq 6.0~\mathrm{GeV}^2$,
which is divided into 18 non-uniformly sized intervals according to the available 
statistics. The interval sizes are at least five times larger than the $Q^2$ 
resolutions, ensuring that the migration effects between neighboring bins are negligible.

For each mode, the signal is modeled by a templated shape derived from 
the corresponding signal MC sample, convolved with a Gaussian function
to account for the mass resolution difference between data and MC simulation. 
The background is described by a second-order polynomial function in mode I 
and an anti-ARGUS~\cite{argus} function in mode II. 
The signal yields from the two modes are constrained to the differential cross section through
\begin{align*}\label{eq:EventsI}
N_{\mathrm{sig},i}^{\mathrm{I(II)}}(Q_{\mathrm{bin},i}^2)
&= L \cdot \mathrm{Br}^{\mathrm{I(II)}} \cdot \varepsilon^{\mathrm{I(II)}}(Q_{\mathrm{bin},i}^2) \nonumber \\
&\quad \cdot \Delta\sigma^{\mathrm{Born}}(Q_{\mathrm{bin},i}^2) / C_{\mathrm{NLO},i},
\end{align*}
where $L$ is the integrated luminosity, $\mathrm{Br}^{\mathrm{I(II)}}$
is the branching fraction of $\eta^\prime$ decays~\cite{PDG} for mode I (II),
$\varepsilon^{\mathrm{I(II)}}(Q_{\mathrm{bin},i}^2)$ is the detection efficiency,
$\Delta\sigma^{\mathrm{Born}}(Q_{\mathrm{bin},i}^2)$ is the Born-level differential
cross section integrated over the $i$-th $Q^2$ bin,
and $C_{\mathrm{NLO},i}$ is the radiative correction factor,
defined as the ratio of the cross section including next-to-leading order (NLO) corrections to
that at the Born level, obtained from the {\sc ekhara} generator~\cite{EKHARA}.
The fit result for $Q^{2}\in [0.55,~0.65]~\mathrm{GeV}^2$ is shown 
in Fig.~\ref{fig:com_all}, and the results for other $Q^2$ bins can be found 
in the supplemental material~\cite{supp}.    
The determined $\Delta \sigma^{\rm Born}(Q^2_{\mathrm{bin},i})$ values and their statistical uncertainties are listed in Table~\ref{tab:Result}.

The squared modulus of the TFF, $|F(Q^2)|^2$, is obtained as
\begin{equation*}
|F(Q^2)|^2 = \frac{\Delta \sigma^{\rm Born}(Q^2)}{\Delta \sigma^{\rm WZW}(Q^2)},
\end{equation*}
where $\Delta \sigma^{\rm WZW}(Q^2)$ is derived
from the point-like Born cross section, $\sigma^{\rm WZW}$, obtained 
with MC simulation of the signal processes using only the Wess-Zumino-Witten
(WZW) term~\cite{WZW1, WZW2} and assuming a constant TFF.
All numerical inputs to the TFF determination, together with 
the product of the average $Q^2$ and the TFF, 
$Q^2|F(Q^2)|$, are summarized in Table~\ref{tab:Result},
and statistical uncertainties are directly derived from the simultaneous fits.
An additional correction of $+1.0\%$ should be applied 
when comparing the measured TFF with calculations at $Q^2_{\rm miss}=0~\mathrm{GeV}^2$.
This correction factor is determined by comparing the TFF obtained from MC simulation based on the triple-octet model with $Q_{\rm miss}^2\in [0,0.08]~\mathrm{GeV}^2$ and the TFF calculated in the same model at $Q_{\rm miss}^2~=0~\mathrm{GeV}^2$. 
For each $Q^2$ interval, the average momentum transfer $\overline{Q^2}$ is evaluated according to~\cite{Q2Average} 
\begin{equation*}
\frac{d\sigma}{d Q^2}(\overline{Q^2})=\frac{1}{\Delta Q^2}\int_{Q^{2}_{\rm low}}^{Q^2_{\rm up}}\frac{d\sigma}{d Q^2}(Q^2)dQ^2,
\end{equation*}
where $Q^{2}_{\rm low}$ and $Q^2_{\rm up}$ are the lower and upper boundaries 
of the corresponding $Q^2$ interval, and $\frac{d\sigma}{d Q^2}(Q^2)$ is the $Q^2$-dependent cross section. 
Replacing the MC cross section used in the calculation with the measured one yields a negligible uncertainty in $\overline{Q^2}$. 
Figure~\ref{fig:sys_all} presents $Q^2|F(Q^2)|$ in comparison 
with previous measurements from CELLO~\cite{ICELLO},
CLEO~\cite{ICLEO}, BaBar~\cite{IBABAR}, and L3~\cite{IL3}. 
This measurement covers $Q^2\in [0.1, ~6.0]~\mathrm{GeV}^2$ with 
a finer binning, provides the first direct determination 
in the region $Q^2 < 0.3~\mathrm{GeV}^2$, 
and achieves a precision better than $3.0\%$ for $Q^2<1.5~\mathrm{GeV}^2$, 
thereby improving upon previous measurements in both accuracy and $Q^2$ coverage. 

\begin{table*}[htp]
  \centering
  \small
  \caption{Numerical inputs for the TFF determination and the product of the weighted average $Q^2$ and the TFF.
  Shown are the $Q^2$ interval, the average $Q^2$, the detection efficiencies for modes I and II $\varepsilon^{\rm I(II)}$,
  the (point-like) Born-level differential cross section integrated over the $Q^2$ interval divided by the interval width
  $\Delta\sigma^{\rm Born}(Q^2)/\Delta Q^2$ ($\Delta\sigma^{\rm WZW}(Q^2)/\Delta Q^2$) and its statistical uncertainty,
  the radiative correction factor $C_{\rm NLO}$, and the product of the weighted average $Q^2$ and the TFF $\overline{Q^2}|F(Q^2)|$. The first uncertainties are statistical, and the second are systematic.
  }
  \label{tab:Result}
  \begin{tabular*}{0.95\linewidth}{@{\extracolsep{\fill}} c c c c c c c c c}
    \hline
    \hline
    $Q^{2}$ interval & $Q^2$ & $\varepsilon^{\rm I}$ & $\varepsilon^{\rm II}$ & $\Delta \sigma^{\rm Born}(Q^2)/\Delta Q^{2}$ & $C_{\rm NLO}$ & $\Delta \sigma^{\rm WZW}(Q^2)/\Delta Q^{2}$ & $|F(Q^2)| \times 10^{-2}$ & $Q^{2}|F(Q^{2})|$ \\  
    $[\mathrm{GeV}^{2}]$ & $[\mathrm{GeV}^{2}]$ & \% & \% & [pb/$\mathrm{GeV}^{2}$] & & [pb] & [$\mathrm{GeV}^{-1}$]& [MeV] \\
    \hline
$0.10-0.25$ & 0.170 & $0.7$  & $0.1$ & $153.4 \pm 6.7$   & $1.025$ & $1582.0 \pm 1.4$ & $31.1 \pm0.7$ & $53.0 \pm1.2 \pm 0.9$ \\
$0.25-0.35$ & 0.299 & $6.0$  & $2.2$   & $50.0 \pm 1.4$  & $1.026$ & $804.3 \pm 0.7$  & $24.8 \pm0.4$ & $74.2 \pm1.1 \pm1.3$ \\
$0.35-0.40$ & 0.375 & $13.3$ & $6.4$   & $34.0 \pm 1.0$  & $1.024$& $612.3 \pm 0.5$  & $23.5 \pm0.4$ & $88.3  \pm1.3 \pm1.3$ \\
$0.40-0.45$ & 0.424 & $19.8$ & $10.1$   & $26.1 \pm 0.7$ & $1.023 $ & $527.0 \pm 0.5$ & $22.2 \pm0.3$ & $94.4  \pm1.3 \pm1.2$ \\
$0.45-0.50$ & 0.476 & $24.2$ & $12.4$  & $21.4 \pm 0.6$  & $1.024 $ & $457.6 \pm 0.4$ & $21.6 \pm0.3$ & $103.0  \pm1.5 \pm1.8$ \\
$0.50-0.55$ & 0.525 & $25.3$ & $12.9$  & $17.2 \pm 0.6$  & $1.022 $ & $402.5 \pm 0.4$  & $20.6 \pm0.3$ & $108.4 \pm1.7 \pm2.3$ \\
$0.55-0.65$ & 0.599 & $25.1$ & $13.0$  & $13.9 \pm 0.3$  & $1.024 $  & $339.4 \pm 0.3$  & $20.3 \pm0.2$ & $121.2 \pm1.5 \pm2.0$ \\
$0.65-0.80$ & 0.721 & $24.5$ & $12.8$  & $8.76 \pm 0.2$  & $1.021 $  & $264.8 \pm 0.2$  & $18.2 \pm0.2$ & $131.1 \pm1.7 \pm1.9$ \\
$0.80-1.00$ & 0.896 & $22.9$ & $11.8$  & $5.29 \pm 0.2$  & $1.019 $  & $197.0 \pm 0.2$  & $16.4 \pm0.2$ & $146.7 \pm2.2 \pm2.3$  \\
$1.00-1.20$ & 1.096 & $26.4$ & $14.9$  & $3.16 \pm 0.1$  & $1.014 $  & $146.8 \pm 0.1$  & $14.7 \pm0.3$ & $160.9 \pm2.8 \pm2.3$  \\
$1.20-1.50$ & 1.344 & $31.5$ & $18.3$  & $1.90 \pm 0.07$  & $1.013 $  & $108.58 \pm 0.10$ & $13.2 \pm0.2$ & $177.6 \pm3.0 \pm3.5$  \\
$1.50-1.80$ & 1.646 & $31.8$ & $19.2$  & $1.06 \pm 0.05$  & $1.011 $  & $78.70 \pm 0.07$  & $11.6 \pm0.3$ & $191.4 \pm4.2 \pm4.1$  \\
$1.80-2.20$ & 1.988 & $31.9$ & $19.6$  & $0.58 \pm 0.03$  & $1.007 $  & $57.36 \pm 0.05$ & $10.0 \pm0.3$ & $199.5 \pm5.3 \pm4.1$  \\
$2.20-2.60$ & 2.389 & $31.6$ & $19.9$  & $0.33 \pm 0.03$  & $1.004 $  & $41.47 \pm 0.04$ & $8.9 \pm0.3$ & $212.1 \pm8.1 \pm3.3$  \\
$2.60-3.00$ & 2.788 & $31.2$ & $19.7$  & $0.20 \pm 0.03$  & $1.000 $  & $31.02 \pm 0.03$ & $8.0 \pm0.4$ & $222.7 \pm10.6 \pm6.5$  \\
$3.00-3.50$ & 3.234 & $30.5$ & $19.5$  & $0.12 \pm 0.01$  & $1.000 $  & $23.39 \pm 0.02$ &$7.2 \pm0.4$ & $233.9 \pm13.4 \pm5.1$  \\
$3.50-4.50$ & 3.961 & $29.8$ & $18.9$  & $0.062 \pm 0.007$ & $0.995 $  & $15.34 \pm 0.01$ &$6.4 \pm0.4$ & $252.1 \pm14.2 \pm6.0$ \\
$4.50-6.00$ & 5.174 & $27.2$ & $17.5$  & $0.016 \pm 0.003$ & $0.995 $  & $8.26 \pm 0.01$ & $4.5 \pm0.4$ & $230.9 \pm22.7 \pm5.0$ \\
    \hline
    \hline
\end{tabular*}
\end{table*}

Systematic uncertainties in the cross section originate mainly from 
the integrated luminosity, the branching fraction of $\eta^\prime$ decays, the detection efficiency, the signal MC model, the radiative correction, and the fit procedure.
Relevant numerical results are presented in the supplementary material~\cite{supp}.
The integrated luminosity is determined from large-angle Bhabha scattering events with an uncertainty of $0.5\%$~\cite{lum0304,lum15}. 
The uncertainties of the $\eta^\prime$ decay branching fractions are taken from the PDG~\cite{PDG}, those for mode I and mode II are $1.2\%$ and $1.3\%$, respectively.

The detection efficiency is influenced by uncertainties in tracking, 
photon reconstruction, PID, the $E/p$ requirement, the requirement 
on $\chi^2$ from the kinematic fit, the angle of the untagged lepton, 
and the limited size of the signal MC sample.
Tracking and PID efficiencies for $\pi^{\pm}$ are studied with 
$D\bar{D}$ events as control samples, while tracking and $E/p$
efficiencies for $e^{\pm}$ are evaluated with radiative Bhabha
events~\cite{Hekk_e}. The associated tracking uncertainties are 
propagated according to the momentum distributions of $\pi^{\pm}$ in the 
signal process and to both the momentum and polar angle distributions 
for $e^{\pm}$.
To improve the precision of the analysis, corrections are applied
to the $\pi^{\pm}$ PID efficiencies and the $E/p$ selection 
efficiencies in the signal MC simulation, the uncertainties of these 
corrections are taken as the systematic uncertainties. 
Photon reconstruction efficiency is studied with radiative 
di-muon events, yielding a $0.5\%$ uncertainty per photon~\cite{PLJ+16}.
The uncertainty from the $\chi^2$ requirement is estimated by 
repeating the analysis without the helix parameter correction 
applied to simulated charged tracks~\cite{Helix}.
The requirement on the angle of the untagged lepton is validated 
using a Barlow test~\cite{Barlow_test}, in which the selection is varied, and the stability of the obtained cross section is checked; additional uncertainties are assigned when the deviation to the central value exceeds two times the uncorrelated uncertainties and exhibits a trend. 
For most $Q^2$ intervals, no additional systematic uncertainty needs to be taken into account. 
The resulting total uncertainties for the detection efficiency range 
from {\color{black}$0.8\%$ and $3.2\%$ for mode I and from $1.5\%$ and $3.6\%$ for mode II, depending on the $Q^2$ interval}. 

The uncertainty due to the model used for signal MC simulation is 
estimated by using an alternative set of parameters in the triple-octet model, where the $\eta-\eta^\prime$ mixing parameters are fixed in the fit.
The resulting change in efficiency is taken as the systematic uncertainty.
The uncertainty from the radiative corrections arises from missing 
higher-order terms beyond NLO. 
Following Ref.~\cite{IBESIII}, we conservatively take $10\%$ for the missing higher-order radiative corrections, which are expected to be of order $(\alpha/\pi)\ln(Q^2/m_e^2)$. 
In addition, the missing higher-order terms may affect detection 
efficiency~\cite{EKHARA}, which is evaluated by comparing MC simulations with and without NLO radiative corrections. 
We take the product of these two factors as the systematic uncertainty from radiative corrections. The uncertainties
increase as $Q^2$ increases, varying from $1.3\%$ to $2.6\%$.

Fit-related uncertainties are evaluated by varying the background description and the fit range. 
The background function is varied from a second- to a third-order polynomial function in mode I, while in mode II, the threshold parameter of the anti-ARGUS function is varied between being free and fixed, the resulting changes in the determined yields are assigned as the systematic uncertainties. 
The robustness of the fit range is studied by performing a Barlow test~\cite{Barlow_test}, in which the fit range is enlarged or shrunk,
and no systematic effect is found. 

Assuming all sources are independent, and accounting for correlations 
between mode I and mode II, the total systematic uncertainty in the 
cross section measurement is determined to range from $2.5\%$ to $4.7\%$, depending on the $Q^2$ intervals. By uncertainty propagation, the uncertainty in the TFF is half of that of the cross section. A summary of the systematic uncertainties can be found in 
supplemental material~\cite{supp}.

In summary, we report a precise measurement of the $\eta^\prime$ TFF in the space-like region with the single-tag method, covering the $Q^2$ region between $0.1~\mathrm{GeV}^2$ and $6.0~\mathrm{GeV}^2$ in 18 bins. 
This measurement fills the gap between the two most precise previous results: BaBar~\cite{IBABAR} at $Q^2$ larger than $4.0~\mathrm{GeV}^2$ and L3~\cite{IL3} at $Q^2$ below $0.15~\mathrm{GeV}^2$. 
In contrast to L3, the present study determines $Q^2$ directly 
from the tagged lepton, providing a direct determination of 
the TFF in the low-$Q^2$ region. 
The $Q^2$ binning chosen for this measurement is substantially finer than that of all previous experiments, enabling a detailed mapping of the TFF evolution and imposing more stringent constraints on theoretical models.

The determined TFF values represent the most precise measurement to 
date and are systematically higher than previous results.
This discrepancy may be related to the larger $Q^2_{\rm miss}$ 
in earlier measurements, for which the assumption of a quasi-real photon becomes less reliable.
With the improved control of $Q^2_{\rm miss}$ in this work, the measured TFF values show good agreement with LQCD prediction~\cite{LQCD2020_Pi0EtaEtap} and holographic model calculation~\cite{Holographic}, while lying above the data-driven phenomenological model calculations~\cite{DataDriven2_EtaEtap, DataDriven3_EtaEtap}.
We note that the contributions to $a_{\mu}^{\eta^\prime}$
predicted by LQCD and the holographic model are larger than those
obtained in the data-driven approach~\cite{DataDriven3_EtaEtap}. 
In light of the new results presented here, existing data-driven 
approaches may therefore require re-evaluations.
The result will also provide constraints for the calculation of the 
$\eta^\prime$ production in gluon-gluon fusion processes~\cite{HHC}.
\\

\begin{acknowledgments}
The BESIII Collaboration thanks the staff of BEPCII (https://cstr.cn/31109.02.BEPC) and the IHEP computing center for their strong support. This work is supported in part by National Key R\&D Program of China under Contracts Nos. 2025YFA1613900, 2023YFA1606000, 2023YFA1606704; National Natural Science Foundation of China (NSFC) under Contracts Nos. 12375070, 11635010, 11935015, 11935016, 11935018, 12025502, 12035009, 12035013, 12061131003, 12192260, 12192261, 12192262, 12192263, 12192264, 12192265, 12221005, 12225509, 12235017, 12342502, 12361141819, 12535005; the Chinese Academy of Sciences (CAS) Large-Scale Scientific Facility Program; the Strategic Priority Research Program of Chinese Academy of Sciences under Contract No. XDA0480600; CAS under Contract No. YSBR-101; 
Shanghai Leading Talent Program of Eastern Talent Plan under Contract No.JLH5913002;
100 Talents Program of CAS; The Institute of Nuclear and Particle Physics (INPAC) and Shanghai Key Laboratory for Particle Physics and Cosmology; Agencia Nacional de Investigación y Desarrollo de Chile (ANID), Chile under Contract No. ANID CCTVal CIA250027; ERC under Contract No. 758462; German Research Foundation DFG under Contract No. FOR5327; Istituto Nazionale di Fisica Nucleare, Italy; Knut and Alice Wallenberg Foundation under Contracts Nos. 2021.0174, 2021.0299, 2023.0315; Ministry of Development of Turkey under Contract No. DPT2006K-120470; National Research Foundation of Korea under Contract No. RS-2026-25486791; National Science and Technology fund of Mongolia; Polish National Science Centre under Contract No. 2024/53/B/ST2/00975; STFC (United Kingdom); Swedish Research Council under Contract No. 2019.04595; U. S. Department of Energy under Contract No. DE-FG02-05ER41374
\end{acknowledgments}

\end{document}